\documentclass[journal]{IEEEtran}

\usepackage{xcolor,soul,framed} %,caption
\usepackage{graphicx}
\usepackage{amsmath}
\usepackage{array}
\usepackage{mdwmath}
\usepackage{mdwtab}
\usepackage{eqparbox}
\usepackage{url}

\begin{document}

\title{Federated Learning and Wireless Communications}

%\title{Federated Learning and Communications \\
%with Wireless Applications}

\author{Zhijin Qin, Geoffrey Ye Li, and Hao Ye \thanks{Zhijin Qin is with Queen Mary University of London, London E1 4NS, UK (e-mail: z.qin@qmul.ac.uk)}
\thanks{Geoffrey Ye Li and Hao Ye are with  Georgia Institute of Technology, Atlanta, GA 30332 USA (e-mail: liye@ece.gatech.edu; yehao@gatech.edu).
}}
\maketitle
\begin{abstract}
Federated learning becomes increasingly attractive in the areas of wireless communications and machine learning due to its powerful functions and potential applications. In contrast to other machine learning tools that require no communication resources, federated learning exploits communications between the central server and the distributed local clients to train and optimize a machine learning model. Therefore, how to efficiently assign limited communication resources to train a federated learning model becomes critical to performance optimization. On the other hand, federated learning, as a brand new tool, can potentially enhance the intelligence of wireless networks. In this article, we  provide a comprehensive overview on the relationship between federated learning and wireless communications, including basic principle of federated learning, efficient communications for training a federated learning model, and federated learning for intelligent wireless applications. We also identify some future research challenges and directions at the end of this article.    
\end{abstract}


\section{Introduction}

With successes of {\it machine learning} (ML), especially {\it deep learning} (DL), in the  areas, such as image recognition and natural language processing, researchers in the communications community have also applied DL to improve the performance of communications or make networks more intelligent recently. DL can improve signal processing performance of communications systems~\cite{DLPhy}. Traditionally, a communication system consists of several modules, such as coding and decoding, modulation and demodulation, channel estimation and signal detection. An intuitive approach is to use a deep neural network (DNN) to represent one or more modules. The whole transmitter or receiver can be even represented by a DNN, which is emerging as end-to-end communications. Of course, we can exploit {\em expert knowledge} in the area of telecommunications accumulated in the past century  to simplify the structure of the DL model and speed up its convergence, which is model-driven DL for physical layer communications.

DL can be also applied in resource allocation in communication networks. Judicious resource allocation can significantly improve the performance of a communication network. Traditionally, resource allocation is formulated as an optimization problem and then  a resource allocation approach can be obtained by solving the corresponding optimization problem. Usually, the optimization problems formulated for wireless resource allocation are {\it non-deterministic polynomial-time} (NP) hard and therefore are quite complicated, if not impossible, to obtain the optimal solutions. As indicated in \cite{DLRA}, DL can  reduce the complexity and improve the performance on solving the optimization problem.  {\it Deep reinforcement learning} (DRL) can be directly exploited in resource allocation, where the environment  contains channel quality, interference level, etc., the action space includes spectrum access, power allocation, spatial resources, etc., and the reward could be composed of  latency, data  rate, etc. Through DRL, a good policy for resource allocation can be obtained to maximize the designed reward. More applications of ML in future wireless networks can be found in~\cite{DLAll}. 

Note that most works on ML for communications are based on centralized learning. Recently, federated learning~\cite{FLGoogle,FLGoogle2} has been proposed to perform model training  distributively at multiple participating clients, each with a part of training data, and coordinated by a central server. By doing so, the computation is offloaded from the central server to the local clients.  Moreover, in federated learning, the participating local clients communicate with the central server only on the model parameters learnt locally rather than the raw data, which preserves privacy in addition to significant reduction on communication overhead. Therefore, federated learning is desired in many privacy-sensitive applications, such as training an imagine classification model based on the photos stored at different mobile devices.

%\textcolor{red}{Depending on scenarios, different names have been used for the central server and the local clients. In mobile edge computing, they are usually called as the remote cloud and edge notes, respectively. Sometimes, they are also referred  to the remote parameter server and workers. While in cellular networks, they could be a base station and mobile devices. }

In brief, federated leaning is enabled by communications between the local clients and the central server and can also be  used as a tool to improve the performance of wireless systems. This article discusses the relationship between federated learning and wireless communications. After introducing the basis of federated learning in Section II, we present efficient communications for federated learning in Section III and federated learning for wireless applications in Section IV. Then, we conclude our article by identifying some research challenges and directions in Section V.   

\section{Federated Leaning}
Even if lots of related work had been done before, the name of {\em federated learning} was first used in~\cite{FLGoogle} in 2016 for collaborative learning in wireless networks, where communication resources, such as bandwidth and transmission power, are limited and the privacy of local clients needs to be preserved. As shown in Figure~\ref{fig1}, we aim to train a global model at the central server with parameters ${\bf w}$ by the whole dataset ${\cal D}=\cup_{k=1}^{K} {\cal D}_k$, where ${\cal D}_k$ is the raw data stored at client $k$. A straightforward way is to let distributed local clients send their data to the central server and then train the  model as in centralized leaning. However,  local clients may be unwilling to share their raw data with others in some privacy-sensitive applications even if they are willing to participate in collaborative model training. Moreover, sending raw data to the central server also consumes significant communication resources, especially when the data size is huge and the number of participating local clients is large. Federated learning  can be adopted to address the above issues. 

%\textcolor{red}{In this section, we will first introduce the basic principle, then present some practical issues on federated learning, and briefly discuss the difference between federated learning and traditional distributed learning.}

\subsection{Basic Principles}

\begin{figure}[!t]
\centerline{\includegraphics[width=3.4in]{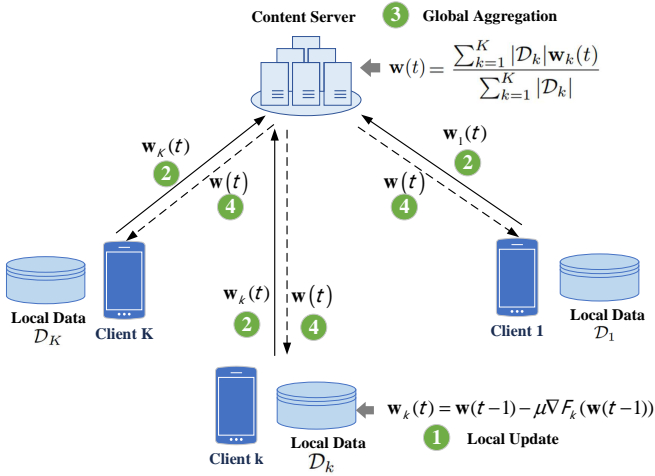}}
\caption{Basic principles of federated learning.}
\label{fig1}
\end{figure}

Following the broad definition in~\cite{FLGoogle2}, federated learning is a machine learning setting where multiple local clients collaborate in training a model under the coordination of a central server while the raw data is kept at the  local clients and the local clients and the central server only communicate on the model parameters. 
As shown in Figure~\ref{fig1}, federated learning include four steps: 
\begin{enumerate}
 \item {\bf Local update}: Each client updates the learning model locally and in parallel according to its raw data;
\item {\bf Weights upload}: Each local client sends its intermediate results, i.e.,  the updated parameters of the trained model $\bf w_k\left(t\right)$,  to the central server; 
\item {\bf Global aggregation}: The central server calculates the average weights, ${\bf{w}}\left(t\right)$, based on parameters received from local clients; 
\item {\bf Weights feedback}: The server broadcasts the updated parameters to each local client for the next iteration. 
\end{enumerate}
Without requiring the transmission of raw data, federated learning addresses the privacy issue, reduces the communication overhead, and offloads the computation from the central server to the local clients. In the following, we introduce the details involved in the aforementioned four steps.

\subsubsection{Loss Function}
The empirical loss function depends on both  model parameters and raw data. Therefore, for raw data, ${\cal D}_k$, stored at  different local clients, the corresponding loss functions is usually different even for the learning model with the same parameter set, ${\bf w}$. Denote $F_k({\bf w})$ as the loss function corresponding to client $k$ with raw data ${\cal D}_k$ and model parameter set ${\bf w}$. Then the global loss function corresponding to the whole dataset, ${\cal D}=\cup_{k=1}^{K} {\cal D}_k$, can be expressed as 
\begin{equation}
    F({\bf w})=\frac{\sum_{k=1}^{K}|{\cal D}_k| F_k({\bf w})}{\sum_{k=1}^{K}|{\cal D}_k|},
\end{equation}
where $|{\cal D}_k|$ denotes the number of elements in   ${\cal D}_k$ at the client $k$, and $|{\cal D}|=\sum_{k=1}^{K} |{\cal D}_k|$.

The model is trained to minimize loss function $F({\bf w})$, for example using the gradient descent approach, to find the optimal parameter set, ${\bf w}^{o}=\arg\min{F({\bf w})}$.
Since raw data is distributed at different clients, we cannot directly find the gradient at the central server as in centralized learning. 

\subsubsection{Model Weights}
If using the gradient descent approach to minimize the global loss function, then we have
\begin{equation}
    {\bf w} (t) :=  {\bf w} (t-1)- \mu \nabla F({\bf w}(t-1)) =  \frac{\sum_{k=1}^{K}|{\cal D}_k| {\bf w}_k (t)}{\sum_{k=1}^{K}|{\cal D}_k|},
\end{equation}
where $\mu$, a small positive number, is the step size, $\nabla F(*)$ represents the gradient of function $F(*)$, 
${\bf w}(t)$ is the global aggregated parameter set at the central server at time $t$, ${\bf w}_k(t)$ is the local parameter set of client $k$ at time $t$, which can be expressed as
\begin{equation}
{\bf w}_k (t) = {\bf w} (t-1)- \mu \nabla F_k({\bf w}(t-1)).
\end{equation}
Afterwards, the central server can calculate ${\bf w} (t)$ as long as the local gradient, $\nabla F_k({\bf w}(t-1))$, is obtained. Therefore, only the local gradients need to be sent to the central server, which  can save communication resource, especially when gradient compression is used as we will discuss in Section III.A. However, we  still have to feed the locally updated parameter set, ${\bf w}_k (t)$, to the central server if the relationship between the updated local parameter set, ${\bf w}_k (t)$, and the previous global parameter set, ${\bf w} (t-1)$, is nonlinear. In this case, the parameter aggregation  could be a more complicated function of ${\bf w}_k (t)$ for $k=1,\cdots, K$ other than that in (2).

\begin{figure}[!t]
\centerline{\includegraphics[width=3.1in]{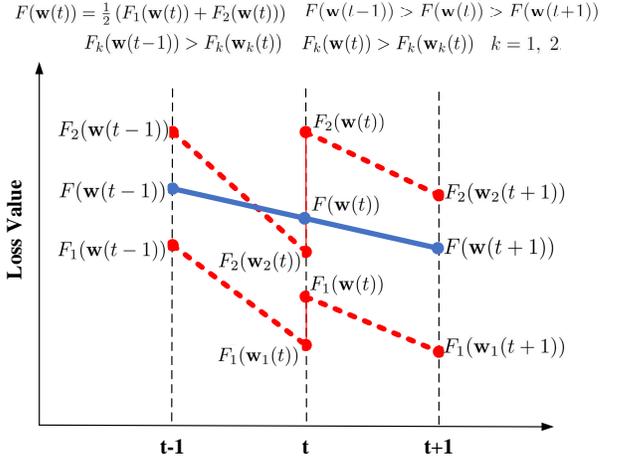}}
\caption{Parameter aggregation and gradient descent optimization.}
\label{fig2}
\end{figure}

It should be emphasized that  (2) ensures that the gradient descent optimization can be performed at the central server through aggregating the gradients of the local loss functions at the distributed clients. Figure~\ref{fig2} demonstrates a simple example for the case with two local clients containing data of the same size. Denote the two local loss functions as $F_1(*)$ and $F_2(*)$ and the global loss function as $F(*)=\frac{1}{2}\left( F_1(*) + F_2(*) \right)$. With gradient descent iteration as in (2), $F({\bf w}(t))$ decreases as $t$ increases, that is, $F({\bf w}(t-1))>F({\bf w}(t))>F({\bf w}(t+1))$. Moreover, the updated local parameter set, ${\bf w}_k(t)$, is obtained through the gradient descent approach corresponding to the local loss function, $F_k(*)$, based on the previous global parameter set, ${\bf w}(t-1)$; therefore, $F_k({\bf w}(t-1))> F_k({\bf w}_k(t))$ for $k=1,~2$. Local parameter sets are then send to the central server for aggregation by ${\bf w}(t)=\frac{1}{2}\left( {\bf w}_1(t) + {\bf w}_2(t) \right)$, which takes the gradients of both local clients into consideration. It is obvious that $F_k({\bf w}(t))>F_k({\bf w}_k(t))$ since ${\bf w}_k(t)$ is obtained based on the gradient descent for the local loss function while ${\bf w}(t)$ is based gradient descent for the global loss function as indicated by (2).

%It should be emphasized that equation (2) ensures that the gradient descent optimization can be performed at the central server through aggregation of the gradients of the local loss functions at distributed local clients. Figure 2 demonstrates a simple example for the case with two local clients of the same size datasets, where the contours of the two local loss functions, $F_1({\bf w})$ and $F_2({\bf w})$, are in red and that of the global loss function, $F({\bf w})=\frac{1}{2} \left( F_1({\bf w}) + F_1({\bf w}) \right)$, is in blue. The red arrows the figure are the gradients of  the two local loss functions, $\nabla F_1({\bf w})$ and $\nabla F_2({\bf w})$, which are different in general and sometimes are even in opposite directions since they corresponding to different local datasets. We can always find the gradient of the global loss function by $\nabla F({\bf w})=\frac{1}{2} \left( \nabla F_1({\bf w}) + \nabla F_2({\bf w}) \right)$, as shown in the black arrow in the figure.

\subsection{Practical Issues}

So far, we have just introduced the basic principle of federated learning. In the following, we identify some issues that cannot be neglected in the design and application of federated learning. 

%For example, it is not clear how to choose the step size for the gradient descent approach for federated learning. If it is too small, the training iteration will converge too slowly and incur too much communication resources. On the other hand, the training iteration will not converge at all if it is too large.
\subsubsection{Stochastic Gradient Descent}
Since loss function $F_k({\bf w}(t))$ is determined by the raw data at the local client $k$,  there is no closed-form expression usually. Therefore, it is impossible to analytically find the gradients, $\nabla F_k({\bf w}(t))$, that is, we cannot carry out one step update as in (3). Usually, {\em stochastic gradient descent} (SGD) is used to find the gradient iteratively at the local clients for each iteration between the central server and  local clients. Readers can refer to \cite{FLGoogle,Comp_Hu} for more details. 

\subsubsection{Robust Aggregation}
For the parameter aggregation in (2), weighted average is used, which only considers the sizes of the raw data and is optimal only if the central server receives local parameter sets {\em accurately} and {\em simultaneously}. In reality, it is difficult to transmit parameter sets in full-precision. Nevertheless, there also exist corruption and noise during parameter exchanges between the central server and the local clients. Robust aggregation in~\cite{Agg_Robust} can address corrupted local parameter sets. As a result of heterogeneous computational capabilities and communication links for different local clients, there may be some local clients, called {\em stragglers}, that deliver their updated local parameter sets later than others and affect timely model parameter updates. The coded federated learning method in~\cite{Code_Intel} may deal with stragglers and speed up the convergence of model training. The broadband analog aggregation method in~\cite{Agg_Huang} considers signal distortion and interference during parameter transmission and results in lowered latency.  

\subsubsection{Upload Frequency}
In the above discussion, we assume that all local clients participate in parameter update at every iteration, which is sometimes inefficient. For example, if the corresponding local parameters update is negligible, then its impact on the aggregation can be ignored. In wireless edge learning, even if the local parameters are non-trivial, the corresponding channel condition could be quite poor. As a result, large amounts of wireless resources, e.g. transmission power and bandwidth, are required to compensate for channel distortion. In this situation, we should jointly consider the occupied communication resources and its contribution to the performance improvement to optimize the whole learning process~\cite{ComDing}.

\subsubsection{Privacy Leakage}
In federated learning, the local clients communicate with the central server on the model parameters, rather than the raw data, to preserve privacy. However, it has been found recently that at least partial privacy information can be recovered from the gradients, which implies that federated learning cannot completely guarantee the privacy of clients. 
\subsubsection{Interplay of Federated Learning and Wireless Communications}
Federated learning depends on wireless communications between the central server and the local clients. Therefore, it is important to efficiently exploit limited wireless resources~\cite{ComGunduz,ComLeung}, which will be discussed in Section III. 

Even if originally proposed to address the concerns on privacy, device computation and storage, and communication bandwidth, federated learning has already had many wireless applications~\cite{FL4Com_Bennis,FL4Com_Bennis2,FL4Com_Cui}, such as wireless resource allocation and localization, as we can see from Section IV.

\subsection{Federated Learning vs Distributed Learning}

Depending on applications and settings, federated learning has different variants, such as {\em cross-device} and {\em cross-silo} federated learning~\cite{FLGoogle2}.  %\textcolor{red}{In the original setting for collaborative learning in mobile devices as in~\cite{FLGoogle}, it is called {\em cross-device federated learning}. It is usually called {\em cross-silo federated learning} for some new application settings, such finance risk prediction and health records mining, where only a small number of reliable clients are involved. THIS CAN BE DELETED} 
Even if not unanimously agreed, some researchers regard federated learning as a kind of distributed learning. 
Nevertheless, the following  prominent characteristics~\cite{FLGoogle2} distinguish federated learning from  traditional distributed learning:
\begin{itemize}
    \item Data is generated locally and remains descentralized at different local clients;
    \item The central server conducts model training, but can never see the raw data at the the local clients;
    \item The local clients only communicate (usually through wireless links) with the central server to update the model parameters.
\end{itemize}
In addition to the above, federated learning also has the following characteristics, for instance, the central server has no control over the local clients and each client can decide whether participating in the collaboration or not; the local clients can be different categories of devices with different distributions on the corresponding data. 

In contrast to federated learning, communication capability in traditional distributed learning is not a bottleneck since the central server and the local clients are usually connected to each other using wirelines or optical fibers, rather than through wireless channels. Most traditional distributed learning approaches have been developed in order to offload computation to the local clients, rather than addressing the privacy or communications issues. Therefore, the whole dataset may be stored at the central server, there could be raw data exchange between the central server and the local clients and among the local clients, or the whole dataset can be even re-partitioned if required. In terms of the distribution scales, there are usually up to 1,000 clients in distributed learning while it could be as large as $10^{10}$ clients for federated learning. 

Nevertheless, we have only noticed the strict definition of federated learning in~\cite{FLGoogle2} so far. To the best of our knowledge, there are no unanimous definitions on distributed learning, descentralized learning, or collaborative learning.

\section{Communications in Federated Learning}

In federated learning, the central server and the local clients keep on exchanging the updated model parameter sets during the training process, which still consumes significant amounts of communication resources, especially when the distributed local clients are wireless connected devices and with a huge number, such as in {\em Internet-of-Things} (IoT) applications. To address the issue, we can compress the information to be transmitted and/or allocate limited communication resource efficiently. Many data compression and communication resource management techniques developed for general purposes can be used in federated learning. Furthermore, if the learning model is a neural network, then we can perform neural network pruning and parameter pruning. In this section, we introduce gradient compression, over-the-air computation, and join optimization techniques, which especially fits  federated learning well. 

\subsection{Gradient Compression}

It has been reported in~\cite{Comp_Hu} that   redundancy in the stochastic gradient could be as large as $99\%$ in certain situations. Therefore, spectrum bandwidth can be significantly saved if the stochastic gradient is  compressed properly. In addition to those general data compression techniques, especially compressive sensing, to address the issue, gradient compression in~\cite{Comp_Hu} exploits the sparsity of stochastic gradient and has been  designed specifically for federated learning. 

Gradient compression includes quantization and sparsification. Gradient quantization converts gradient elements, which are usually continuous, into (low-precision) discrete values to facilitate digital transmission. A simple way is to quantify the original stochastic gradient into binary. Improved versions include three-level and four-level quantization. Gradient sparsification refers to removing the gradient elements with small amplitudes. An intuitive way is to turn the elements with amplitudes below a threshold into zero. However, it is sometimes hard to choose the threshold. Another way is to turn off a certain number of elements with small amplitudes into zero. It has been demonstrated that, by properly combining gradient quantization and sparsification, the compression ratio can reach as small as $2.5\%$ in certain situations while reasonable convergence performance of model training is still maintained. 

{\em Deep gradient compression} (DGC) proposed in~\cite{Comp_Hu} further improves the performance of compression by the following four steps: momentum correction, local gradient clipping, momentum factor masking, and warm-up training. DGC can achieve a compression radio of $0.17\%$ in certain situations.

We should emphasize that the gradient compression performance strongly depends on the learning models. Most models currently studied are for image recognition and language processing. Therefore, it is not clear whether these results still hold for wireless applications. 

\subsection{Over-the-Air Computation}

If the multiple local clients transmit their updated local parameter sets through the same wireless channel simultaneously, the received signal at the central server will be the superposition of all the local parameter sets. Surprisingly, the addition or weighted average is computed over the air.  On the other hand,
the aggregated global parameter set depends only on the weighted average of the local parameter sets as in (2). Inspired by this observation, a group of works exploit the {\em over-the-air computation} property of the wireless multiple-access channel to obtain the weighted average, that is, to perform aggregation, directly at the central server, without requiring the individual local parameter set. In that case, significant communication bandwidth can be saved, especially when there are a massive number of local clients, as in IoT applications.

\begin{figure}[!t]
\centerline{\includegraphics[width=3.4in]{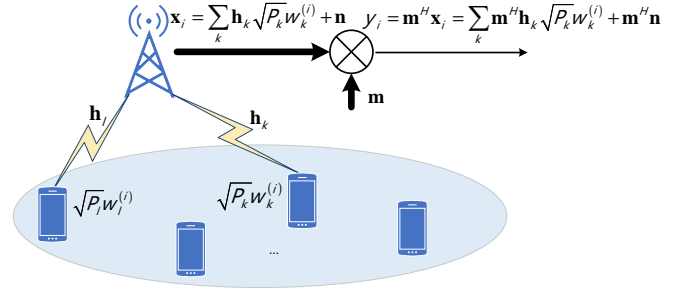}}
\caption{Parameter aggregation through over-the-air computation and exploiting spatial freedom.}
\label{fig3}
\end{figure}

If there are multiple antennas at the receiver of the central server, there will be more freedom to assign spatial resource and optimize federated learning. Figure~\ref{fig3} demonstrates the principle of the method in~\cite{ComDing}. The local clients simultaneously send their $i$-th elements of the local parameter sets with proper power scales, $\sqrt{p_k}w_k^{(i)}$ for $k=1, \cdots K$, through the wireless channel. Then the received signal vector at the central server will be ${\bf x}_i= \sum_k {\bf h}_k \sqrt{p_k} w_k^{(i)}$ +{\bf n}, where ${\bf h}_k$ and ${\bf n}$ are the channel gain vector corresponding to local client $k$ and the noise vector, respectively. After  combining with beamforming vector ${\bf m}$, we have $y_i={\bf m}^T{\bf x}_i=\sum_k {\bf m}^T {\bf h}_k \sqrt{p_k} w_k^{(i)} + {\bf m}^T{\bf n}$. The beamforming vector here provides freedom for efficient transmission. If there were no noise or maximum power constraint for each local client, then ${\bf m}$ and $p_k$ should be selected such that ${\bf m}^T{\bf h}_k \sqrt{p_k}=\frac{|{\cal D}_k|}{\sum_l |{\cal D}_l|}$ for perfect aggregation according to equation (2). With channel distortion and a maximum transmission power constraint for each local client in reality, there may be no enough freedom for perfect aggregation, especially when there are extensive number of local clients. In~\cite{ComDing}, a sparse and low-rank optimization problem is formulated to address the issue.  The effective approach in~\cite{ComDing} selects a subset of local clients with proper transmission power and carefully adjusts the beamforming vector to optimize the statistical performance of federated learning.  

\subsection{Joint Local Compression and Global Aggregation}
\begin{figure}[!t]
\centerline{\includegraphics[width=3.5in]{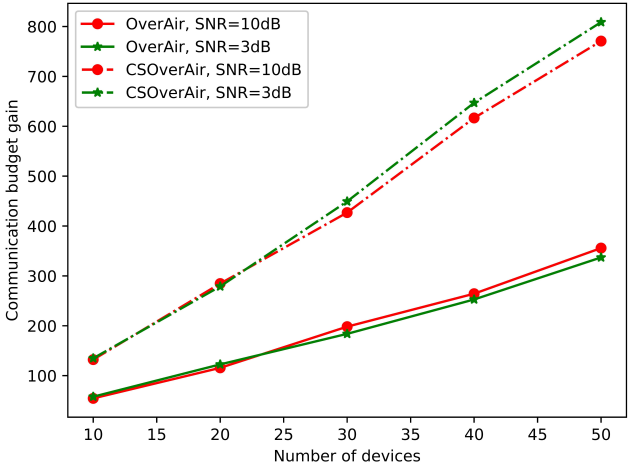}}
\caption{Communication budget gain of the over-the-air computation with parameter compression at the local clients and recovery/aggregation at the central server.}
\label{fig4}
\end{figure}

In~\cite{ComGunduz}, all local clients compress their gradients by compressive sensing and then send to the central server through the same wireless channel simultaneously, the central server then reconstructs and aggregates the gradient information from the noisy observation, which combines gradient compression and over-the-air computation.  The {\it CSOverAir} in~\cite{ComGunduz}  sparsifies the gradients by a compression ratio of 1\% at each local client.   Figure~\ref{fig4} compares the communication  gain of the {\it over-the-air computation approach (OverAir)} and the CSOverAir, where the communication gain is measured as the ratio of the communication overhead of the scheme without over-the-air computation or gradients compression  and that of the OverAir and CSOverAir. From the figure,  the communication overhead is reduced significantly by over-the-air computation as multiple clients could transmit their parameters to the center server simultaneously by using the same channel. It is also noted that the communication overhead can be reduced greatly by compressing the gradients at each local client. However, the minimal compression ratio in the CSOverAir approach~\cite{ComGunduz} is heavily dependent on the compressive sensing techniques, which fails to consider the features of chanenl fading  during the recovery at the central server.

The frequency of global parameter aggregation is linked to the communication overhead and the convergence speed of model training. In~\cite{ComLeung}, the convergence of gradient-descent based federated learning is analyzed and a convergence bound is obtained. Then, a control algorithm is developed based on the convergence bound. After learning data distribution, system dynamics, and model characteristics, the control algorithm adjusts the frequency of aggregation to minimize the global loss function with given communication resources. To efficiently exploit limited communication resources, joint optimization of parameter compression, communication resource allocation, and model training/aggregation can be performed, which is also a promising direction for further research.

\section{Federated Learning for Wireless Applications}

In the above, we have introduced the principle of federated learning and efficient communication for distributed model training. Federated learning also has many wireless applications. In this section, we will provide three examples: vehicular communications~\cite{FL4Com_Bennis}, {\em unmanned aerial vehicle} (UAV) online path control~\cite{FL4Com_Bennis2}, and localization~\cite{FL4Com_Cui}.  

\subsection{Distribution Estimation for Vehicular Communications}

Vehicular communications will make our daily vehicular operation safer, greener, and more efficient and it also paves the path to intelligent deriving. In addition to communicate with the infrastructure (V2X), vehicles also need to exchange critical information, such as safety related messages, with their surrounding vehicles (V2V) with ultra reliability and low latency, which is referred to as {\em ultra-reliable low-latency communication} (URLLC). 

To ensure URLLC by optimally allocate limited communication resources, i.e., frequency bands and transmission power,
we have to model and capture extremely low probability events, such as the probability of the queue length over a threshold or extreme high reliability of a wireless link, and find their relationship with the assigned communication resources. Considering high dynamics and high mobility of vehicular networks, it is often impossible to find an exact or closed-form expression. Fortunately, {\em extreme value theory} (EVT) can address the issue~\cite{FL4Com_Bennis}, which models a rare event by the generalized Pareto distribution determined by several critical parameters and converts to the issue into estimating these critical parameters. 

Federated learning has been used in~\cite{FL4Com_Bennis} to estimate these critical parameters and catch the relationship between assigned resources, link reliability, and transmission latency. Different from the classical {\em maximum likelihood estimation} (MLE) that needs to transmit data from different vehicle users to the roadside unite, the critical parameters can be obtained using federated learning based MLE, which can save communication resources for data transmission and avoid privacy issues. It has been demonstrated in~\cite{FL4Com_Bennis} that the federated learning based estimation can save up to $79\%$ data exchange but with estimation accuracy similar to the central solution.  

\subsection{Massive UAV Online Path Control}

\begin{figure}[!t]
\centerline{\includegraphics[width=3.3in]{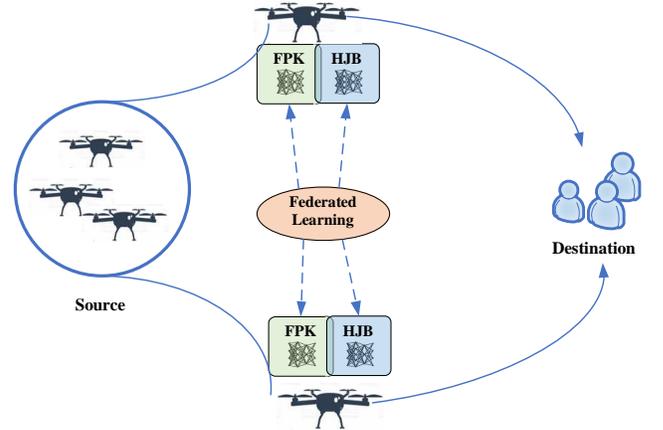}}
\caption{Massive UAV control based on mean-field game and federated learning. }
\label{fig5}
\end{figure}

Sometimes, a large population of UAVs may fly from a source to a destination for mission-critical tasks, such as helping firefighting or covering disaster areas as temporal access points. Due to wind and other random factors, massive UAV control becomes very challenging in order to avoid collisions and reach the destination quickly as in Figure~\ref{fig5} from~\cite{FL4Com_Bennis2}. To limit inter-communications among UAVs, mean-field game (MFG) is used for massive UAV control by solving a pair of coupled stochastic differential equations: the {\em Fokker-Plank-Kolmogorov} (FPK) equation to estimate the population distribution and the {\em Hamilton-Jacobi-Bellman} (HJB) equation to control the UAV flight, i.e., direction, speed, and acceleration~\cite{FL4Com_Bennis2}. As shown in Figure~\ref{fig5}, two DNNs at each UAV are used to approximate the solutions of the FPK and HJB equations. To ensure the convergence and accelerate the DNN training, federated learning is leveraged by periodically exchanging the parameters sets of the DNNs for solving the FPK and the HJB equations. From~\cite{FL4Com_Bennis2}, MFG control based on federated learning can shorten the traveling distance and reduce the probability of collision both by $50\%$. 
\subsection{Distributed Localization}

For a given environment, radio features of a mobile device can uniquely determine its location, which makes it possible to perform localization based on radio features. However, the relationship between the mobile location and the corresponding radio features is usually very complicated. A deep learning model can be used to map the radio features into a specific location. To train the deep learning model, a huge training data on radio features corresponding to different locations is required, which is sometimes challenging. One way is to collect the data on radio features and mobile locations from all mobiles in a certain area and then use them to train the deep learning model for localization, which, however, will cause privacy issues and huge communication overhead. Therefore, federated learning is used in~\cite{FL4Com_Cui} to training machine learning models for localization, which is called {\em fed}erated {\em loc}alization (FEDLOC).

The framework of FEDLOC is similar to generic federated learning in Figure~\ref{fig1}. Each mobile, as a local client, collects local data on radio features and locations, update the model parameter set locally, and send it to the central server. The base station or just a fusion center, as the central server, aggregates the received local parameter sets to obtain the global one. After comparing the two machine learning models for localization, it is found in~\cite{FL4Com_Cui} that the {\em Gaussian processes} model with maximum likelihood loss function is better than the DNN model with least-square loss function based on the testing by real data. 
  
\section{Conclusion and Remarks}

In this article, we have presented the principle and efficient communications of federated learning. We have also discussed some wireless application examples, which reflect the power of federated learning for future wireless systems. However, many issues and challenges remain unexplored. Here we identify some of them. 
\begin{itemize}
\item {\em Over-the-air computation}: It has been proposed for joint uplink communication and aggregation \cite{ComDing,ComGunduz}, where perfect synchronization and accurate channel state information are assumed. In reality, exact synchronization is impossible. Furthermore, wireless channel may have multi-path delay spread. It is desired to address these deployment imperfections on over-the-air computation for model parameter aggregation.
\item {\em Local update and global aggregation}: Instead of sharing all the weights with the server, each local client can keep its own model parameters, which are close to but not the same with the weights of the global model at the server. Models at local clients can  be tailored for their own purposes. Therefore, a key problem is how to update the local models with the broadcast information from the server and how to process and transmit the gradients of the local model to help the training of the global model at the server. One possible solution is using a meta-learner for learning the gradients for local and global models.
\item {\em Privacy}: The privacy of federated learning has been improved significantly comparing with  sending the raw data to the central server straightforwardly, there is still partial privacy leakage as we have indicated in Section II.B. It is still desired to further improve the privacy in federated learning. 

\item {\em New wireless applications}: A wireless network can be regarded as a distributed learning system, which fits in with applications of federated learning well. In addition to distributed localization~\cite{FL4Com_Cui},  more wireless applications are expected, such as mobile edge caching and resource allocation in vehicular networks.
\end{itemize}


\begin{thebibliography}{}
\bibitem{DLPhy} Z.-J.~Qin, H.~Ye, G.~Y.~Li, and B.-H.~Juang, “Deep learning in physical layer communications,” {\em IEEE Wireless Commun.}, vol.~26, no.~2, pp.~93-99, Apr. 2019. 

\bibitem{DLRA}L.~Liang, H.~Ye, G.-D.~Yu, and G.~Y.~Li, “Deep learning based wireless resource allocation with application in vehicular networks,” {\em Proc. IEEE} vol.~108, no.~2, pp.~341 - 356, Feb.~2020.  

\bibitem{DLAll} C.-X.~Jiang, H.-J.~Zhang, Y.~Ren, Z.~Han, K.-C.~Chen, and L.~Hanzo, “Machine learning paradigms for next-generation wireless networks,” {\em IEEE Wireless Commun.}, vol.~24, no.~2, pp.~98-105, Apr. 2017. 

\bibitem{FLGoogle} H.~B.~McMahan, E.~Moore, D.~Ramage, S.~Hampson, and B.~A.~Arcas, “Communication-efficient learning of deep networks from descentralized data,” at https://arxiv.org/abs/1602.05629v3, Feb.~2017.

\bibitem{FLGoogle2} P.~Kairouz, etc., “Advances and open problem in federated learning,” at https://arxiv.org/abs/1912.04977v1, Dec.~2019.

%\bibitem{FMTL} V.~Smith, V.-K.~Chiang, M.~Sanjabi, and A.~Talwalkar, “Federated multi-task learning,” at https://arxiv.org/abs/1705.10467v2, Feb.~2018.

\bibitem{Comp_Hu} Y.-J. Lin, S.~Han, H.-Z. Mao, Y.~Wang, and W.~J.~Dally, “Deep gradient compression: Reducing the communication bandwidth for distributed training,” in {\em Proc. ICLR} 2018.

\bibitem{Agg_Robust} K.~Pillutla, S.~M.~Kakade, and Z.~Harchaoui, "Robust aggregation for federated learning," https://arxiv.org/abs/1912.13445v1, Feb.~2020.

\bibitem{Code_Intel} S.~Dhakal, S.~Prakash, Y.~Yona, S.~Talwar, and N.~Himayat, “Coded federated learning,” at https://arxiv.org/abs/2002.09574v1, Feb.~2020.

\bibitem{Agg_Huang} G.-X.~Zhu, Y.~Wang, and K.-B.~Huang, "Broadband analog aggregation for low-latency federated edge learning," {\em IEEE Trans.~Wireless Commun.}, vol.~19, no.~1, Jan.~2020.

\bibitem{ComDing} K.~Yang, T.~Jiang, Y.-M.~Shi, and Z.~Ding, “Federated learning via over-the-air computation,” at https://arxiv.org/abs/1812.11750v3, Feb.~2019.

\bibitem{ComGunduz} M.~Mohammadi and D.~G\"{u}nd\"{u}z,  “Machine learning at the wireless edge: distributed stochastic gradient descent over-the-air,” at https://arxiv.org/abs/1901.00844v2, Feb.~2019.

\bibitem{ComLeung} S.-Q.~Wang, T.~Tuor, T.~Salonidis, K.~K.~Leung, C.~Makaya, T.~He, and K.~Chan, “Adaptive federated learning in resource constrained edge computing systems,” {\em IEEE J.~Select.~Areas Commun.}, vol.~37, no.~6, pp.~1205-1221, Jun. 2019.

\bibitem{FL4Com_Bennis} S.~Samarakoon, M.~Bennis, W.~Saad, and M.~Debbah, "Distributed federated learning for ultra-reliable low-latency vehicular communications," {\em IEEE Trans.~Commun.}, vol.~68, no.~2, pp.~1146-1159, Feb.~2020.

\bibitem{FL4Com_Bennis2} H.~Shiri, J.-H.~Park and M.~Bennis, "Communication-efficient massive UAV online path control: federated learning meets mean-field game theory," at https://arxiv.org/abs/2003.04451v1, Mar. 2020.

\bibitem{FL4Com_Cui} F.~Yin, Z.-D.~Lin, Y.~Xu, Q.-L.~Kong, D.~S.~Li, S.~Theodoridis, and S.-G. R. Cui, "FEDLOC: Federated learning framework for cooperative localization and location data processing," at https://arxiv.org/abs/2003.03697v1, Mar. 2020.



\end{thebibliography}
\end{document}